\def\unit{\relax{\rm 1\kern-.29em l}}
\def\alb{{\bar \alpha}}
\def\beb{{\bar \beta}}
\def\gab{{\bar \gamma}}
\def\deb{{\bar \delta}}
\def\rhb{{\bar \rho}}
\def\lab{{\bar \lambda}}
\def\ab{{\bar a}}

\def\jb {\bar{j}}
\def\kb {\bar{k}}
\def\lb {\bar{l}}
\def\mb {\bar{m}}
\def\nb {\bar{n}}
\def\pb {\bar{p}}
\def\qb {\bar{q}}

\def\sb {\bar{s}}

\def\wb {\bar{w}}
\def\ub {\bar{u}}
\def\vb {\bar{v}}
\def\zb {\bar{z}}

\def\Nb {\bar{N}}

\input lanlmac
\input epsf

\noblackbox

%% Macro
%style

%general

%Macro for figure
\newcount\figno
\figno=0
\def\fig#1#2#3{
\par\begingroup\parindent=0pt\leftskip=1cm\rightskip=1cm\parindent=0pt
\baselineskip=11pt
\global\advance\figno by 1
\midinsert
\epsfxsize=#3
\centerline{\epsfbox{#2}}
\vskip 12pt
{\it Figure \the\figno:} #1\par
\endinsert\endgroup\par
}
\def\figlabel#1{\xdef#1{\the\figno}}

%%%%% NEW DEF OOKOUCHI%%%%%%%%
\def\le{\leq}
\def\ge{\geq}

%%%%%%%%%%%%%%%%%%%%%%%%%%%%%%%

%%%%%%%%%%%%%%%%%%%%%%%%%%%%%%%

% References
%\ColemanPY
%\AharonyPA
\lref\AharonyPA{
  O.~Aharony, M.~Berkooz and E.~Silverstein,
  ``Multiple-trace operators and non-local string theories,''
  JHEP {\bf 0108}, 006 (2001)
  [arXiv:hep-th/0105309].
  %%CITATION = JHEPA,0108,006;%%
}

%\CachazoJY
\lref\CachazoJY{
  F.~Cachazo, K.~A.~Intriligator and C.~Vafa,
  ``A large $N$ duality via a geometric transition,''
  Nucl.\ Phys.\  B {\bf 603}, 3 (2001)
  [arXiv:hep-th/0103067].
  %%CITATION = NUPHA,B603,3;%%
}

%\KachruIH
\lref\KachruIH{
  S.~Kachru, S.~H.~Katz, A.~E.~Lawrence and J.~McGreevy,
  ``Open string instantons and superpotentials,''
  Phys.\ Rev.\  D {\bf 62}, 026001 (2000)
  [arXiv:hep-th/9912151].
  %%CITATION = PHRVA,D62,026001;%%
}

\lref\ColemanPY{
  S.~R.~Coleman,
  ``The fate of the false vacuum. 1. semiclassical theory,''
  Phys.\ Rev.\ D {\bf 15}, 2929 (1977)
  [Erratum-ibid.\ D {\bf 16}, 1248 (1977)].
  %%CITATION = PHRVA,D15,2929;%%
  %%Cited 833 times in SPIRES-HEP
}

%\DuncanAI
\lref\DuncanAI{
  M.~J.~Duncan and L.~G.~Jensen,
  ``Exact tunneling solutions in scalar field theory,''
  Phys.\ Lett.\ B {\bf 291}, 109 (1992).
  %%CITATION = PHLTA,B291,109;%%
}

%\ArgyresXH
\lref\ArgyresXH{
  P.~C.~Argyres and A.~E.~Faraggi,
  ``The vacuum structure and spectrum of ${\cal N}=2$ supersymmetric $SU(n)$ gauge
  theory,''
  Phys.\ Rev.\ Lett.\  {\bf 74}, 3931 (1995)
  [arXiv:hep-th/9411057].
  %%CITATION = PRLTA,74,3931;%%
}

%\KlemmQS
\lref\KlemmQS{
  A.~Klemm, W.~Lerche, S.~Yankielowicz and S.~Theisen,
  ``Simple singularities and ${\cal N}=2$ supersymmetric Yang-Mills theory,''
  Phys.\ Lett.\  B {\bf 344}, 169 (1995)
  [arXiv:hep-th/9411048].
  %%CITATION = PHLTA,B344,169;%%
}

%\DouglasNW
\lref\DouglasNW{
  M.~R.~Douglas and S.~H.~Shenker,
  ``Dynamics of $SU(N)$ supersymmetric gauge theory,''
  Nucl.\ Phys.\  B {\bf 447}, 271 (1995)
  [arXiv:hep-th/9503163].
  %%CITATION = NUPHA,B447,271;%%
}

%\AlberghiTU
\lref\AlberghiTU{
  G.~L.~Alberghi, S.~Corley and D.~A.~Lowe,
  ``Moduli space metric of ${\cal N} = 2$ supersymmetric $SU(N)$ gauge theory and the
  enhancon,''
  Nucl.\ Phys.\  B {\bf 635}, 57 (2002)
  [arXiv:hep-th/0204050].
  %%CITATION = NUPHA,B635,57;%%
}

%\IntriligatorDD
\lref\IntriligatorDD{
  K.~Intriligator, N.~Seiberg and D.~Shih,
  ``Dynamical SUSY breaking in meta-stable vacua,''
  JHEP {\bf 0604}, 021 (2006)
  [arXiv:hep-th/0602239].
  %%CITATION = JHEPA,0604,021;%%
}

%\OoguriPJ
\lref\OoguriPJ{
  H.~Ooguri and Y.~Ookouchi,
  ``Landscape of supersymmetry breaking vacua in geometrically realized gauge
  theories,''
  Nucl.\ Phys.\  B {\bf 755}, 239 (2006)
  [arXiv:hep-th/0606061].
  %%CITATION = NUPHA,B755,239;%%
}

%\FrancoES
\lref\FrancoES{
  S.~Franco and A.~M.~Uranga,
  ``Dynamical SUSY breaking at meta-stable minima from D-branes at obstructed
  geometries,''
  JHEP {\bf 0606}, 031 (2006)
  [arXiv:hep-th/0604136].
  %%CITATION = JHEPA,0606,031;%%
}

%\KitanoXG
\lref\KitanoXG{
  R.~Kitano, H.~Ooguri and Y.~Ookouchi,
  ``Direct mediation of meta-stable supersymmetry breaking,''
  Phys.\ Rev.\  D {\bf 75}, 045022 (2007)
  [arXiv:hep-ph/0612139].
  %%CITATION = PHRVA,D75,045022;%%
}

%\OoguriBG
\lref\OoguriBG{
  H.~Ooguri and Y.~Ookouchi,
  ``Meta-stable supersymmetry breaking vacua on intersecting branes,''
  Phys.\ Lett.\  B {\bf 641}, 323 (2006)
  [arXiv:hep-th/0607183].
  %%CITATION = PHLTA,B641,323;%%
}

%\ArgurioNY
\lref\ArgurioNY{
  R.~Argurio, M.~Bertolini, S.~Franco and S.~Kachru,
  ``Gauge/gravity duality and meta-stable dynamical supersymmetry breaking,''
  JHEP {\bf 0701}, 083 (2007)
  [arXiv:hep-th/0610212].
  %%CITATION = JHEPA,0701,083;%%
}

%\HigashijimaWZ
\lref\HigashijimaWZ{
  K.~Higashijima and M.~Nitta,
  ``Kaehler normal coordinate expansion in supersymmetric theories,''
  Prog.\ Theor.\ Phys.\  {\bf 105}, 243 (2001)
  [arXiv:hep-th/0006027].
  %%CITATION = PTPKA,105,243;%%
}

%\AlvarezGaumeHN
\lref\AlvarezGaumeHN{
  L.~Alvarez-Gaume, D.~Z.~Freedman and S.~Mukhi,
  ``The background field method and the ultraviolet structure of the
  supersymmetric nonlinear sigma model,''
  Annals Phys.\  {\bf 134}, 85 (1981).
  %%CITATION = APNYA,134,85;%%
}

%\SeibergRS
\lref\SeibergRS{
  N.~Seiberg and E.~Witten,
  ``Electric - magnetic duality, monopole condensation, and confinement in ${\cal N}=2$
  supersymmetric Yang-Mills theory,''
  Nucl.\ Phys.\  B {\bf 426}, 19 (1994)
  [Erratum-ibid.\  B {\bf 430}, 485 (1994)]
  [arXiv:hep-th/9407087].
  %%CITATION = NUPHA,B426,19;%%
}

%\SeibergAJ
\lref\SeibergAJ{
  N.~Seiberg and E.~Witten,
  ``Monopoles, duality and chiral symmetry breaking in ${\cal N}=2$ supersymmetric
  QCD,''
  Nucl.\ Phys.\  B {\bf 431}, 484 (1994)
  [arXiv:hep-th/9408099].
  %%CITATION = NUPHA,B431,484;%%
}

%\SeibergUR
\lref\SeibergUR{
  N.~Seiberg,
  ``Supersymmetry and nonperturbative beta functions,''
  Phys.\ Lett.\  B {\bf 206}, 75 (1988).
  %%CITATION = PHLTA,B206,75;%%
}

%\DineXT
\lref\Dine{
  M.~Dine and J.~Mason,
  ``Gauge mediation in metastable vacua,''
  arXiv:hep-ph/0611312.
  %%CITATION = HEP-PH/0611312;%%
}

%\KitanoXG
\lref\KOO{
  R.~Kitano, H.~Ooguri and Y.~Ookouchi,
  ``Direct mediation of meta-stable supersymmetry breaking,''
  Phys.\ Rev.\  D {\bf 75}, 045022 (2007)
  [arXiv:hep-ph/0612139].
  %%CITATION = PHRVA,D75,045022;%%
}

%\MurayamaYF
\lref\MurayamaI{
  H.~Murayama and Y.~Nomura,
  ``Gauge mediation simplified,''
  arXiv:hep-ph/0612186.
  %%CITATION = HEP-PH/0612186;%%
}

%\CsakiWI
\lref\Terning{
  C.~Csaki, Y.~Shirman and J.~Terning,
  ``A simple model of low-scale direct gauge mediation,''
  arXiv:hep-ph/0612241.
  %%CITATION = HEP-PH/0612241;%%
}

%\ArgurioQK
\lref\ArgurioII{
  R.~Argurio, M.~Bertolini, S.~Franco and S.~Kachru,
  ``Metastable vacua and D-branes at the conifold,''
  arXiv:hep-th/0703236.
  %%CITATION = HEP-TH/0703236;%%
}

%\FrancoHT
\lref\FrancoII{
  S.~Franco, I.~Garcia-Etxebarria and A.~M.~Uranga,
  ``Non-supersymmetric meta-stable vacua from brane configurations,''
  JHEP {\bf 0701}, 085 (2007)
  [arXiv:hep-th/0607218].
  %%CITATION = JHEPA,0701,085;%%
}

%\BenaRG
\lref\Bena{
  I.~Bena, E.~Gorbatov, S.~Hellerman, N.~Seiberg and D.~Shih,
  ``A note on (meta)stable brane configurations in MQCD,''
  JHEP {\bf 0611}, 088 (2006)
  [arXiv:hep-th/0608157].
  %%CITATION = JHEPA,0611,088;%%
}

%\KawanoRU
\lref\KaOO{
  T.~Kawano, H.~Ooguri and Y.~Ookouchi,
  ``Gauge mediation in string theory,''
  arXiv:0704.1085 [hep-th].
  %%CITATION = ARXIV:0704.1085;%%
}

%\KitanoWM
\lref\KitanoWM{
  R.~Kitano,
  ``Dynamical GUT breaking and mu-term driven supersymmetry
breaking,''
  Phys.\ Rev.\  D {\bf 74}, 115002 (2006)
  [arXiv:hep-ph/0606129].
  %%CITATION = PHRVA,D74,115002;%%
}%\IntriligatorPY

\lref\IntriligatorPY{
  K.~Intriligator, N.~Seiberg and D.~Shih,
  ``Supersymmetry breaking, R-symmetry breaking and metastable
vacua,''
  arXiv:hep-th/0703281.
  %%CITATION = HEP-TH/0703281;%%
}

%\GiveonFK
\lref\GiveonFK{
  A.~Giveon and D.~Kutasov,
  ``Gauge symmetry and supersymmetry breaking from intersecting
branes,''
  arXiv:hep-th/0703135.
  %%CITATION = HEP-TH/0703135;%%
}

%\CachazoSG
\lref\CachazoSG{
  F.~Cachazo, B.~Fiol, K.~A.~Intriligator, S.~Katz and C.~Vafa,
  ``A geometric unification of dualities,''
  Nucl.\ Phys.\  B {\bf 628}, 3 (2002)
  [arXiv:hep-th/0110028].
  %%CITATION = NUPHA,B628,3;%%
}

%\SeibergPQ
\lref\SeibergPQ{
  N.~Seiberg,
  ``Electric - magnetic duality in supersymmetric nonAbelian gauge
theories,''
  Nucl.\ Phys.\  B {\bf 435}, 129 (1995)
  [arXiv:hep-th/9411149].
  %%CITATION = NUPHA,B435,129;%%
}

%\BanksMA
\lref\BanksMA{
  T.~Banks,
  ``Remodeling the pentagon after the events of 2/23/06,''
  arXiv:hep-ph/0606313.
  %%CITATION = HEP-PH/0606313;%%
}

%\CachazoRY
\lref\CachazoRY{
  F.~Cachazo, M.~R.~Douglas, N.~Seiberg and E.~Witten,
  ``Chiral rings and anomalies in supersymmetric gauge theory,''
  JHEP {\bf 0212}, 071 (2002)
  [arXiv:hep-th/0211170].
  %%CITATION = JHEPA,0212,071;%%
}

%% the end of references

%%%%%%%%%%%%%%%%%%%%%%%%%%%%%%%%%%%%%%%%%%%%%%%%%%%%%%%%%%%%%%%%%
%                      Title Page                               %
%%%%%%%%%%%%%%%%%%%%%%%%%%%%%%%%%%%%%%%%%%%%%%%%%%%%%%%%%%%%%%%%%

\newbox\tmpbox\setbox\tmpbox\hbox{\abstractfont }
\Title{\vbox{\baselineskip12pt \hbox{CALT-68-2646}\hbox{UT-07-14}}
}
{\vbox{\centerline{Metastable Vacua}\smallskip\centerline{in Perturbed
Seiberg-Witten Theories}}}
\vskip 0.2cm

\centerline{Hirosi Ooguri,$^{1,2}$ Yutaka Ookouchi$^1$ and Chang-Soon Park$^1$}
\vskip 0.4cm
\centerline{$^1$\it California Institute of Technology, Pasadena,
CA 91125, USA}
\vskip 0.2cm
\centerline{$^2$\it
Department of Physics, University of Tokyo,
Tokyo 113-0033, Japan}

\vskip 1.3cm

%%%%%%%%%%%%%%%%%%%%%%%%%%%%%%%%%%%%%%%%%%%%%%%%%%%%%%%%%%%%%%%%%%%%%%%%%%%%%
\centerline{\bf Abstract}

We show that, for a generic choice of a point on the
Coulomb branch of any ${\cal N}=2$ supersymmetric gauge theory,
it is possible to find a superpotential perturbation which
generates a metastable vacuum at the point. For theories with
$SU(N)$ gauge group, such a superpotential can be expressed
as a sum of single-trace terms for $N=2$ and $3$. If the
metastable point is chosen at the origin of the moduli space,
we can show that the superpotential can be a single-trace
operator for any $N$. In both cases, the superpotential
is a polynomial of degree $3N$ of the vector multiplet
scalar field.

\medskip

%%%%%%%%%%%%%%%%%%%%%%%%%%%%%%%%%%%%%%%%%%%%%%%%%%%%%%%%%%%%%%%%%%%%%%%%%%%%%
\noindent

\bigskip
\bigskip
%\draft
\Date{April 2007} %replace this line by \draft  for preliminary versions
	     %or specify \draftmode at some point

%\vskip 1.3cm

%%%%%%%%%%%%%%%%%%%%%%%%%%%%%%%%%%%%%%%%%%%%%%%%%%%%%%%%%%%%%%%%%%%%%%%%%%%%%
% Abstract                                                                  %
%%%%%%%%%%%%%%%%%%%%%%%%%%%%%%%%%%%%%%%%%%%%%%%%%%%%%%%%%%%%%%%%%%%%%%%%%%%%%
%\noindent

%\bigskip\bigskip
%\Date{April 05, 2007}

%%%%%%%%%%%%%%%%%%%%%%%%%%%%%%%%%%%%%%%%%%%%%%%%%%%%%%%%%%%%%%%%%%
%                        CONTENT                                 %
%%%%%%%%%%%%%%%%%%%%%%%%%%%%%%%%%%%%%%%%%%%%%%%%%%%%%%%%%%%%%%%%%%

%%%%%%%%%%%%%%%%%%%%%%%%%%%%%%%%%%%%%%%%%%%%%%%%%%%%%%%%%%%%%%%%%%%%%%%%%%%%%
\newsec{Introduction}
%%%%%%%%%%%%%%%%%%%%%%%%%%%%%%%%%%%%%%%%%%%%%%%%%%%%%%%%%%%%%%%%%%%%%%%%%%%%%

Since the discovery of metastable vacua in massive SQCD in \IntriligatorDD, supersymmetry breaking at metastable vacua has attracted wide attention. Following their idea and techniques, various phenomenological models have been proposed
\refs{\BanksMA\KitanoWM\Dine\KOO\MurayamaI\Terning-\IntriligatorPY}.
Moreover, one can construct supersymmetry breaking models in the context of string theories as a low energy theories on D-branes
\refs{\FrancoES\OoguriPJ\OoguriBG\FrancoII\Bena\ArgurioNY\ArgurioII\KaOO-\GiveonFK}. These works are done in free magnetic ranges \SeibergPQ  , where metastable
vacua can be found by perturbative analysis.

In this paper, we will take a different route and study the Coulomb branch of ${\cal N}=2$ \ supersymmetric gauge theory \refs{\SeibergRS \SeibergAJ}, perturbed by a small superpotential. As pointed out in \IntriligatorDD, an addition of a mass term $W=m \phi^2$ for the
vector multiplet scalar field $\phi$
does not lead to metastable vacua; only saddle points occur.
We show that metastable vacua can be generated
by choosing a more general form of superpotential,
at almost any point on the moduli space
in any ${\cal N}=2$ supersymmetric gauge theory.
This follows from the fact that the sectional curvature of the Coulomb branch moduli space
is positive semi-definite. This provides another indication for the ubiquity of
metastable vacua in supersymmetric gauge theories.

Gauge theories realized in string theory often have superpotentials which have only
single-trace terms.
For theories with $SU(N)$ gauge group, a single-trace
superpotential can generate a metastable vacuum at any point
in the Coulomb branch when $N=2$ and $3$. We also study
the case when the metastable
point is chosen at the origin of the moduli space. In this case, we find that the
superpotential can be a single-trace operator for any $N$.
In both cases, the superpotential is a polynomial of degree $3N$ in terms of the
vector multiplet scalar field $\phi$.

This paper is organized as follows. In section 2, we show that a metastable vacuum can occur at a generic
point on the Coulomb branch with an appropriate choice of a superpotential.
After developing the general framework,
we discuss the case of the $SU(2)$ theory without matter multiplet \SeibergRS\
in detail to show how the mechanism works. In this case, the potential can be drawn as
a three-dimensional graph, where we can see how a metastable vacuum is
generated explicitly. In section 3, we estimate the life-time of such metastable vacua.

In the appendices, we study several examples explicitly. In Appendix A, we study the moduli stabilization
in the semi-classical regime. In Appendix B, we study the case with $SU(N)$ gauge group \refs{\KlemmQS, \ArgyresXH}.
Although the expression for the metric is complicated for general $N$,
we are able to compute its curvature at the origin of the moduli space.
Using this, we find an explicit form of the superpotential that generates a metastable vacuum
at the origin. In this case, the superpotential can be chosen as a single-trace operator.
Explicitly, in terms of the gauge invariant operators $u_r={\rm tr}(\phi^r)$, the superpotential
$${
W=\lambda \left({1\over N} u_N + {(N-1)^2\over 6N^3} {1\over \Lambda^{2N}} u_{3N}\right)
}$$
for small coupling constant $\lambda$ produces a metastable vacuum at the origin of the moduli space for any $N$, where $\Lambda$ is the scale of the gauge theory.

%%%%%%%%%%%%%%%%%%%%%%%%%%%%%%%%%%%%%%%%%%%%%%%%%%%%%%%%%%%%%%%%%%%%%%%%%%%%%
\newsec{General consideration}
%%%%%%%%%%%%%%%%%%%%%%%%%%%%%%%%%%%%%%%%%%%%%%%%%%%%%%%%%%%%%%%%%%%%%%%%%%%%%

In this section, we show how to construct metastable vacua in the Coulomb branch
of an arbitrary ${\cal N}=2$ supersymmetric gauge theory with gauge group $G$,
possibly with hypermultiplets, by introducing a small superpotential.
The key property of ${\cal N}=2$ gauge theory is that the metric for the moduli space is
(the rigid limit of) special K\"ahler. The effective Lagrangian at the Coulomb branch is generically ${\cal N}=2$ \ $U(1)^{{\rm rank}\ G}$ supersymmetric gauge theory and is described by
\eqn\eLeffSemi{{\cal L}_{eff}={\rm Im} {1 \over 4\pi} \left[ \int d^4 \theta \partial_i {\cal F}(A) \bar A^i + {1\over2} \int d^2 \theta {\partial^2 {\cal F}(A)\over \partial A^i \partial A^j} W^i_{\alpha} W^j_{\alpha}\right]\;,}
where $i,j=1,\cdots, {{\rm rank}\ G}$. It follows that
the metric on the moduli space ${\cal M}$ is given by
\eqn\eSKM{
g_{i\jb} = {\rm Im} \tau_{i j} = {\rm Im} {\partial^2 {\cal F}(a) \over \partial a^i
\partial a^j}\;.
}
Later in this section, we will show that this relation implies that any sectional curvature
of the curvature operator $R$ is positive semi-definite. That is, for any given holomorphic
vector field $w \in T{\cal M}$,
$${
\left<w,R(v,v)w \right> \ge 0 \qquad {\rm for \ all} \; v \in
T{\cal M}_p ~~~ {\rm and \ all} \; p \in {\cal M}\;.
}$$
We call such curvature operator semi-positive.\foot{That the Ricci curvature of the Coulomb
branch is positive semi-definite was noted in \ref\OoguriIN{
  H.~Ooguri and C.~Vafa,
  ``On the geometry of the string landscape and the swampland,''
  Nucl.\ Phys.\  B {\bf 766}, 21 (2007)
  [arXiv:hep-th/0605264].
  %%CITATION = NUPHA,B766,21;%%
}. Here we are making a stronger statement that the sectional curvatures are positive
semi-definite.}
  The curvature is called positive
if the equality holds only when $v=w=0$. In our case, the tensor $\left<w,R(\cdot,\cdot)w\right>$
is strictly positive definite at almost every point on the moduli space.

For a generic point in the moduli space where the curvature is positive, we can show that
a suitable superpotential exists that generates a metastable vacuum at the point.
Of course, the superpotential has to be small so that it does not
affect the K\"ahler potential significantly.
Suppose we parameterize the moduli space using some coordinate system $x^i$ ($i=2,3,\cdots,N$) near a point $p$. We may introduce the K\"ahler normal coordinates $z^i$ \refs{\AlvarezGaumeHN,\HigashijimaWZ} as
\eqn\ezTrans{
z^i=x'^i + {1 \over 2} \Gamma^i_{jk} x'^j x'^k + {1\over 6} g^{i \mb} \partial_l (g_{n\mb}\Gamma^n_{jk}) x'^j x'^k x'^l,
}
where connections are evaluated at $p$ and $x'=x-x(p)$.
The expansion is terminated at the cubic order since higher order terms are
not relevant for our purpose.
 Then the metric in the $z$ coordinate system is
$${g_{i\jb}(z,\zb)=\tilde g_{i \jb} + \tilde R_{i\jb k \lb} z^k \zb^{\lb}+O(z^3)\;,}$$
where \ $\widetilde{}$\ \ \ means evaluation at $p$.
The inverse metric is given by
\eqn\eInvm{g^{i\jb}(z,\zb)=\tilde g^{i \jb} + {\tilde {R}^{i\jb}} _{\,\,\,k \lb} z^k \zb^{\lb}+O(z^3)\;.}
Let us consider a superpotential $W=k_i z^i$. Note that there are global coordinates for the moduli space.
For example, $u_r={\rm tr} (\phi^r)$ are global coordinates in ${\cal N}=2$ $SU(N)$ gauge theory,
and we can write down $W$ in terms of $u_r$ by coordinate transformation. The corresponding
superpotential is then expressed by replacing $u_r$ with ${\rm tr} (\phi^r)$.

Suppose $k_i$ is so small that corrections to the K\"ahler potential
is negligible. Then the leading potential is given by
\eqn\eVI{V=g^{i \bar j} k_i \bar{k}_{\jb} + k_i \bar k_{\bar j}
{\tilde {R}^{i\jb}}_{\,\,\,k \lb} z^k \zb^{\lb}  +O(z^3)\;.}
If $\tilde{R}$ is positive, the potential indeed gives a metastable vacuum at $p$. If $\tilde{R}$ is semi-positive, there could be some flat directions. However,
 if $k_i$ is not along the null direction of $\tilde{R}$, and the tensor $k_i \kb_{\jb}\tilde{R}^{ij}_{\,\,\,k\lb}$ has positive-definite
eigenvalues, we get a metastable vacuum. Generically, these conditions can be satisfied.
For example, in the semi-classical region of the ${\cal N}=2$ $SU(N)$ gauge theory
without hypermultiplets, which we study in Appendix A, we can make metastable
vacua at any point.
On the other hand, in some other examples studied in Appendices, there
arise some flat directions in $k_i$ because we choose a highly symmetric point,
which is not sufficiently generic. Even in these cases, we can find a superpotential to
generate a metastable vacuum by choosing
$k_i$ appropriately.

Now, let us prove the assertion that the curvature $R$ is semi-positive.
Since we are interested in the local behavior, we can use $a^i$ $(i=2,3,\cdots, N)$ as coordinates in which the metric is given by
$${
g_{i\jb} = {\rm Im} \tau_{i j} = {\rm Im} {\partial^2 {\cal F} \over \partial a^i \partial a^j}\;.
}$$
In ${\cal N}=2$ \ $SU(N)$ supersymmetric gauge theory, these $a^i$ are the periods of a meromorphic one-form describing the Coulomb branch. An important fact is that each $\tau_{i j}$ is holomorphic.
In components, we want to show
\eqn\eWTS{
w^j \wb^{\mb} g_{\mb i} R^i_{\,jk\lb} v^k \vb^{\lb} \ge 0 \qquad {\rm for \ all}\quad v,w.
}
Since $\tau_{ij}$ is holomorphic,
\eqn\eR{\eqalign{
R^i_{\,jk\lb}&=-\partial_{\lb} \left( g^{\qb i} \partial_k g_{j\qb}\right) \cr
&=-(\partial_{\lb} g^{\qb i}) \partial_k g_{j\qb} \cr
&=g^{\qb p} g^{i\nb} \partial_{\lb} g_{p \nb} \partial_{k} g_{j\qb}\;.
}}
Plugging this into the LHS of \eWTS,
\eqn\ePositiveDefinite{\eqalign{
w^j \wb^{\mb} g_{\mb i} g^{\qb p} g^{i\nb} \partial_{\lb} g_{p \nb} \partial_{k} g_{j\qb} v^k \vb^{\lb} &= w^j \wb^{\nb}g^{\qb p} \partial_{\lb} g_{p \nb} \partial_{k} g_{j\qb} v^k \vb^{\lb}\cr
&=g^{\qb p} (w^j v^k \partial_k g_{j\qb})(\wb^{\nb} \vb^{\lb} \partial_{\lb} g_{\nb p}) \ge 0
}}
since $g^{\qb p}$ is positive definite. Therefore, \eWTS\ is satisfied.
For a given holomorphic vector field $w$, ${P_{k\qb}}=w^j \partial_k g_{j\qb}$ is holomorphic, so its determinant is 0 only on a complex co-dimension one subspace of the moduli space unless it is a constant. Thus, generically ${P_{k\qb}v^k}$ is nonzero for nonzero $v$, which implies \ePositiveDefinite\ is strictly positive for any nonzero $v$. We found that the curvature is semi-positive and that the tensor $w^j \wb^{\mb} g_{\mb i} R^i_{\,jk\lb}$ is strictly positive definite
at almost every point on the moduli space.

The superpotential $W=k_i z^i$ can be expressed in terms of
 global coordinates of the moduli space, such as
$u_r={\tr (\phi^r)}$ for $SU(N)$, by coordinate
transformation near the metastable vacuum.
Generally terms quadratic and cubic order in $u_r$'s are needed
(higher order terms are not relevant for the metastability), and
the superpotential would contain multiple-trace operators.
On the other hand, gauge theories realized in string theory
often have superpotentials consisting of
single-trace terms only \refs{\CachazoJY,
 \KachruIH, \CachazoSG}.\foot{For discussion of theories with
multiple-trace superpotentials, see \AharonyPA.}
To see when the superpotential can be chosen as a sum of single-trace terms,  let us
consider ${\cal N}=2$ $SU(N)$ gauge theory. For $SU(2)$, the situation is easy since
any multiple-trace operator
can be expressed in terms of a single-trace operator.
This is not the case when the gauge group is $SU(3)$.
However, in this case, we can show that the superpotential
$W = k_i z^i$ can be deformed in such a way that
$W$ turns into a single-trace operator
without destabilizing the metastable point given by the bosonic potential $V = g^{i\bar j}
\partial_i W \bar \partial_{\bar j} \bar W$. To see this,
let $u={\rm tr} \phi^2$ and $v={\rm tr} \phi^3$ be the two coordinates for
$SU(3)$ and let $u'=u-u_0$ and $v'=v-v_0$ be the coordinates centered at
$(u_0,v_0)$. We can express $u_i={\rm tr}\phi^i$ ($i=0,2,3,\cdots,9$) as
polynomials of $u'$ and $v'$. They are all independent generically. To construct
the superpotential that generates a metastable vacuum at $u'=v'=0$,
we can ignore terms that are quartic and higher order in $u'$ and $v'$.
Hence $u_i$ span a 9 dimensional subspace of the 10 dimensional cubic
polynomial space. But the missing polynomial can be set to vanish by
using deformation analogous to the one used in Appendix B.1,
which does not disturb the metastability.
For higher $N$,
we have not been able to find out whether it is possible to construct a single-trace superpotential that can
generate a metastable vacuum at a generic point in the moduli space.
But at the origin of the moduli space, for any $SU(N)$, the single-trace superpotential
$${
W=\lambda \left({1\over N} u_N + {(N-1)^2\over 6N^3} {1\over \Lambda^{2N}} u_{3N}\right)
}$$
for small coupling constant $\lambda$ produces a metastable vacuum , where $\Lambda$ is the scale of the gauge theory, as we show in Appendix B.

%%%%%%%%%%%%%%%%%%%%%%%%%%%%%%%%%%%%%%%%%%%%%%%%%%%%%%%%%%%%%%%%%%%%%%%%%%%%%
\subsec{$SU(2)$ Seiberg-Witten theory }
%%%%%%%%%%%%%%%%%%%%%%%%%%%%%%%%%%%%%%%%%%%%%%%%%%%%%%%%%%%%%%%%%%%%%%%%%%%%%

We can apply our mechanism to produce metastable vacua at strong coupling regime.
Let us demonstrate this at the origin in pure ${\cal N}=2$ $SU(2)$ gauge theory \SeibergRS.
We first construct an appropriate superpotential using the K\"ahler normal coordinate near the origin of the moduli space. Since all expressions for the periods and metric are given in terms of the hypergeometric functions explicitly, we can easily determine the effective potential produced by superpotential perturbation.
Let $u={\rm tr} \phi^2$ be the modulus of the theory. The elliptic curve that describes the moduli space of the $SU(2)$ Seiberg-Witten theory is
$${y^2=(x^2-u)^2 - \Lambda^4\;.}$$
The periods of the theory are given by
$${\eqalign{{\partial a \over \partial u} &= {\sqrt 2 \over 2} (e_2-e_1)^{-1\over 2}(e_4-e_3)^{-1\over 2} F \left({1\over2},{1\over 2}, 1, z\right)\cr
{\partial a_D \over \partial u} &= {\sqrt 2 \over 2} \left[(e_1-e_2)(e_4-e_3) \right]^{-1\over 2} F \left({1\over2},{1\over 2}, 1, 1-z\right)\;,
}}$$
where
$${z={(e_1-e_4)(e_3-e_2)\over(e_2-e_1)(e_4-e_3)}}$$
and
$${
e_1=-\sqrt{u-\Lambda^2},\quad e_2=\sqrt{u-\Lambda^2} ,\quad e_3=\sqrt{u+\Lambda^2} ,\quad e_4=-\sqrt{u+\Lambda^2}\;.
}$$
The periods determine the metric in the $a$ coordinate by
$${\tau={\partial a_D / \partial u \over \partial a / \partial u}}\;.$$
We are going to use the metric in the $u$ coordinate. This can be expanded near the origin:
$${\eqalign{g_{u\ub} &= {\rm Im} \tau \left| {da\over du} \right|^2 = r(1+su^2 +\sb\ub^2 - t u \ub) +O(u^3)}\;,}$$
where $r=0.174 \Lambda^{-2}$, $s=0.125\Lambda^{-4}$ and $t=0.0522\Lambda^{-4}$.

We can use the K\"ahler normal coordinate $z$ given by \ezTrans\ to choose a superpotential
$${W=m z, \qquad  z= u+ {1\over3} s u^3}$$
for small real coupling constant $m$. The corresponding effective potential is
$${V={m^2\over g_{z\zb}}={m^2\over g_{u\ub}} \left|1+su^2\right|^2\;.}$$
The graphs for the potential are drawn in Figure 1 in two different scales. Although the metastable vacuum is visible when magnified near the origin, it can hardly be seen at the scale of the graph on the right. The potential is almost flat near the origin, and the
metastable vacuum is generated by a tiny dip!
Interestingly, this is not due to some small parameters
of the theory. Actually, other than the scale $\Lambda$,
there are no additional parameters that we can put in the theory
if we consider a metastable vacuum at the origin. The near-flatness
of the potential around the origin is generated without fine-tuning.
\bigskip
\centerline{\epsfxsize=0.5\hsize\epsfbox{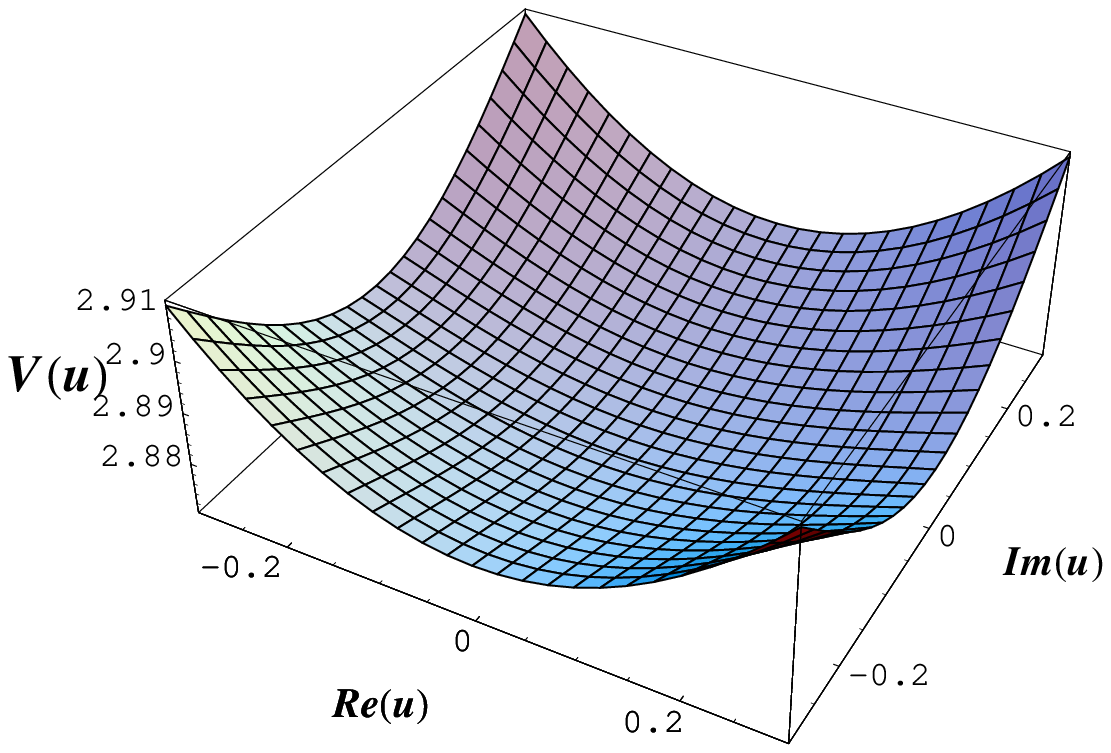}\epsfxsize=0.5\hsize\epsfbox{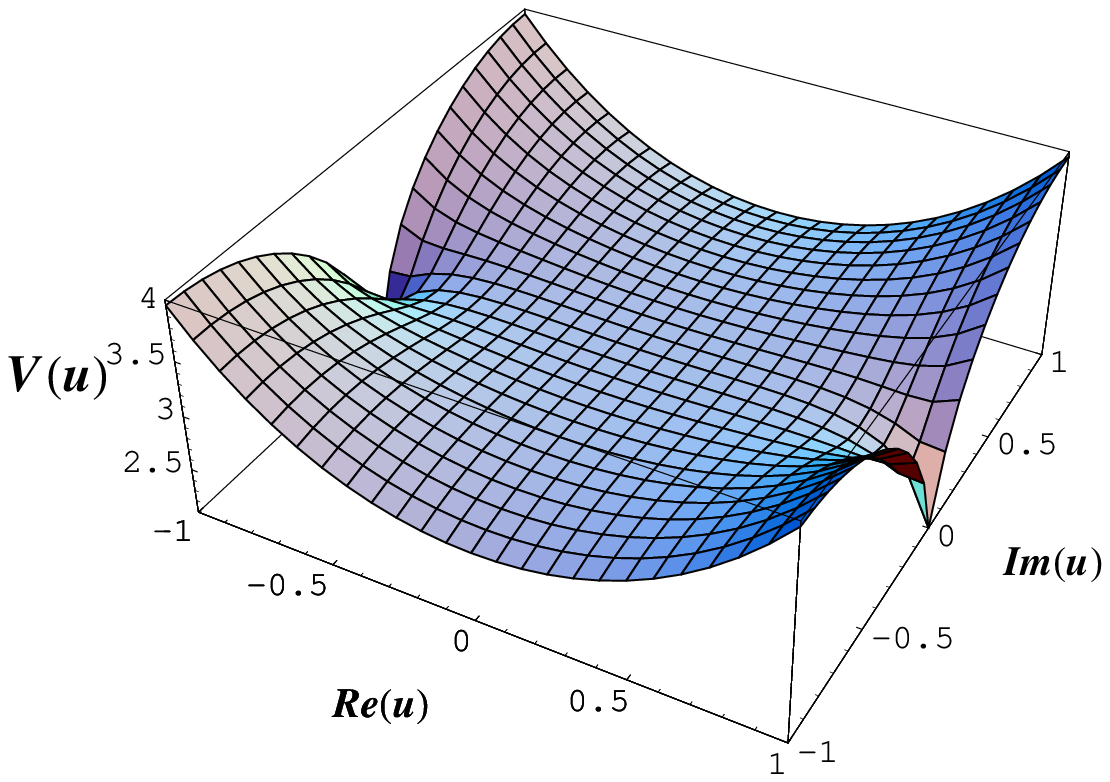}}
%\noindent{\ninepoint\sl \baselineskip=8pt {\bf Figure 1}:{\sl $\;$
%{\bf Figure 1}
\noindent{\ninepoint\sl \baselineskip=8pt {\bf Figure 1}:{\sl
The appearance of a metastable vacuum at the origin. Note difference of scales.}}
\bigskip

We can consider a more general superpotential
\eqn\eGS{ W= mz + {1\over 2} \alpha z^2 + {1\over 3} \beta z^3 \;.}
We have to set $\alpha=0$ to have a local minimum at $z=0$. Using \eInvm , the effective potential $V=g^{z\zb} \left| \partial W/\partial z\right|^2$ becomes
$${ V= m^2 R z\zb + m g^{u\ub} \beta zz + m g^{u\ub} \beb \zb\zb + {\rm constant} + O(z^3)\;,}$$
where $R=R^{z\zb}_{\,\,\,\,\,z\zb} = R^{u\ub}_{\,\,\,\,\,u\ub}=-g^{u\ub}\partial_{\ub} \partial_u \log g_{u\ub} =0.150 \Lambda^{-2}$.
In this case, it is straightforward to read the range to have a local minimum at $z=0$. We need
$${ m R\pm 2 g^{u\ub}\beta >0.}$$
Hence $\left| {\beta\over m} \right| < {g_{u\ub} R\over2}=0.0261\Lambda^{-4}$. In the $u$ coordinate system, \eGS\ becomes
$${W=m \left[u+ {1\over 3} \left(s+{\beta\over m}\right) u^3\right] +O(u^4)\;.}$$
So, we want $s+\beta/m$ to lie between $(0.125\pm0.0261) \Lambda^{-4}$. We can confirm numerically that, precisely in this range, do we have a metastable vacuum at the origin.

We can consider also a superpotential that makes a metastable vacuum at some point other than the origin. This is possible for any points because the curvature is positive everywhere except at the two singular points, where it diverges.
Also, for $SU(2)$ case, any polynomial of $u$ can be expressed as single-trace form.

Now that we have found a metastable vacuum, we want to check its longevity. Notice that the
$SU(2)$ Seiberg-Witten theory has only one dimensionful parameter $\Lambda$. In Figure 1, we have set it to be 1. If we change this, the coordinate $u$ in the graph scales. Therefore, by sending the scale $\Lambda$ to some limit, we may have a long-lived metastable vacuum at the origin. To see this, consider slices of the potential around the origin. Cutting through the real and imaginary axes, the potential looks like Figure 2:
\bigskip
\centerline{\epsfxsize=0.8\hsize\epsfbox{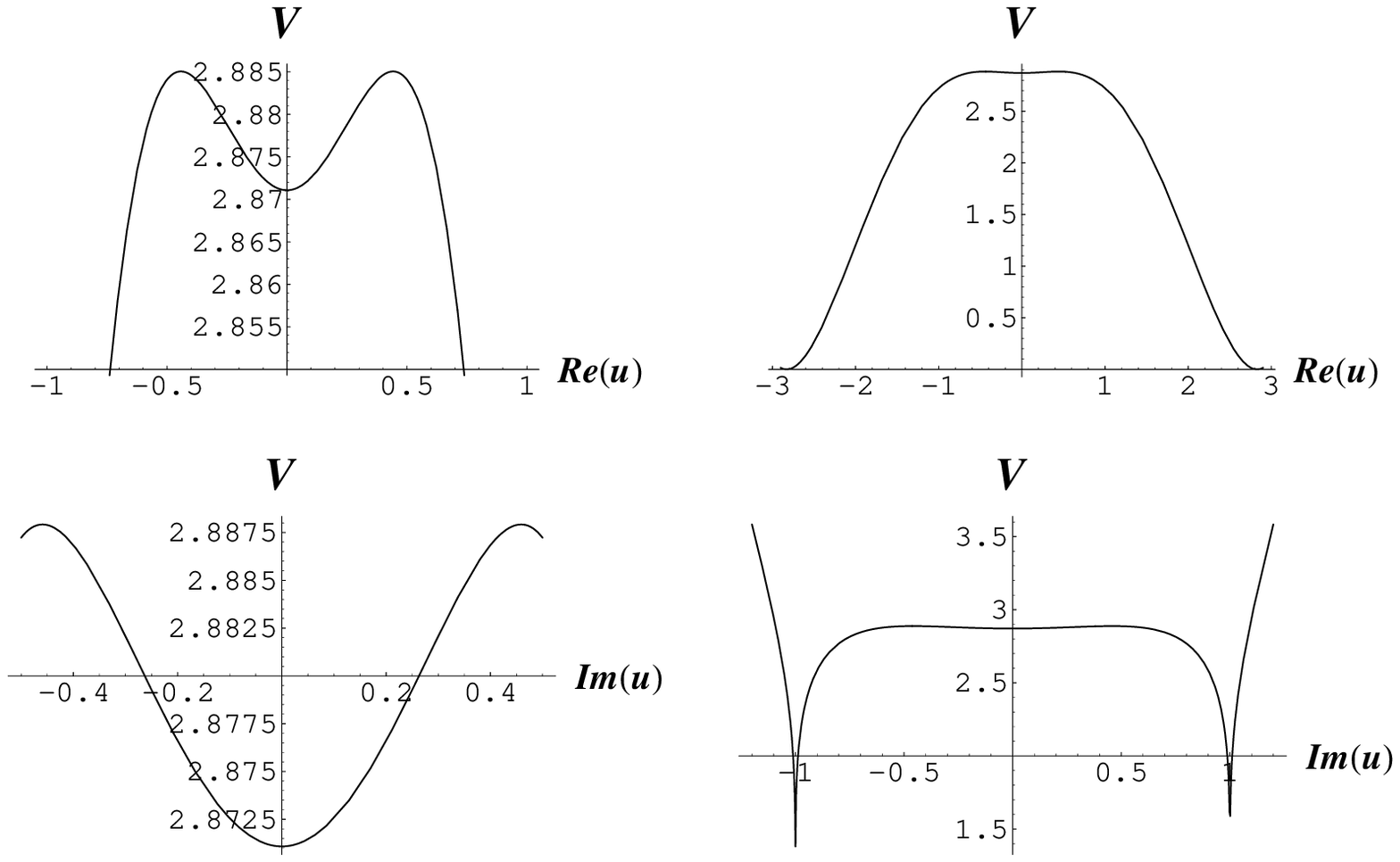}}
\noindent{\ninepoint\sl \baselineskip=8pt {\bf Figure 2}:{\sl
The real and imaginary slices of the potential through the origin, shown in two different scales.}}
\bigskip

We see that the characteristic feature of the graph is that it gets really flattened near the origin, and the local minimum at the origin and the peak of the graph are almost of the same height. But the distance between the origin, where the metastable vacuum is located,
and the supersymmetric vacua can be arbitrarily large by setting $\Lambda$ large. In such a case, we use the triangular approximation \DuncanAI\ instead of the thin-wall approximation \ColemanPY . The tunneling rate is proportional to $e^{-S}$ where
\eqn\eTunnel{
S \sim {(\Delta u / \Lambda)^4\over V_{+}}\;,
}
where $\Delta u$ is the distance between the peak and the origin and, and $V_{+}$ is the difference of the potentials between at the peak and at the origin. We insert $\Lambda$ to make the $u$ field of dimension 1. $\Delta u$ is proportional to $\Lambda^2$. $V_{+}$ is proportional to the mass parameter $m$ in \eGS . Therefore, we can make the bounce action arbitrarily large: we choose $m$ and $\Lambda$ such that $m/\Lambda \ll 1$. This limit accords with our assumption that we have added a small ${\cal N}=2$-to-${\cal N}=1$ supersymmetry breaking term.

Since the superpotential $W=m\left(u+{1\over 3} s u^3\right)$ has a cubic interaction, it introduces supersymmetric vacua when $u=u_0=\pm \sqrt{-1/s}$. We have to consider the tunneling rate to decay into those vacua. However, the distance from 0 to $u_0$ is also set by the scale $\Lambda$. Therefore, for sufficiently large $\Lambda$, the decay process is arbitrarily suppressed.

%%%%%%%%%%%%%%%%%%%%%%%%%%%%%%%%%%%%%%%%%%%%%%%%%%%%%%%%%%%%%%%%%%%%%%%%%%%%%
\newsec{Decay rate of metastable vacua}
%%%%%%%%%%%%%%%%%%%%%%%%%%%%%%%%%%%%%%%%%%%%%%%%%%%%%%%%%%%%%%%%%%%%%%%%%%%%%
In the previous subsection, we considered the decay rate of the metastable vacuum at the origin of the moduli space of the $SU(2)$ Seiberg-Witten theory. Extending the idea, let us estimate the decay rate of metastable vacua constructed using the curvature for a
general ${\cal N}=2$ theory. We do not have an explicit expression for the effective potential. However, we can make a general argument that metastable vacua can be arbitrarily long-lived by choosing parameters appropriately. Note that whenever there appears a massless monopole or dyon in the moduli space, the metric diverges. In such a case, the effective potential vanishes and we get a supersymmetric vacuum at that point. The set of supersymmetric vacua is a sub-variety of the moduli space. Additionally, the superpotentials introduce more supersymmetric vacua. Therefore, it is difficult to compute the exact tunneling rate. But we can estimate its dependence on the scale $\Lambda$ and the typical scale of $k_i$. We consider the most efficient path to go from the metastable vacuum to a supersymmetric one. We expect that the shapes of such 1-dimensional slices enable us to use the triangular approximation \DuncanAI  , just as in $SU(2)$ case. \eTunnel\ in this case becomes
\eqn\eTunnel{
S \sim {(\Delta Z)^4\over V_{+}}\;.
}
Here $\Delta Z$ is the distance between the metastable and supersymmetric vacua in $z$ coordinates, scaled by some power of $\Lambda$ to have a mass dimension 1, and $V_{+}$ is the difference of the effective potentials between at the metastable and supersymmetric vacua. Since the metric of the moduli space is determined by one dimensionful parameter $\Lambda$, $(\Delta Z)^4$ is proportional to $\Lambda^4$. If the coordinates $x$ in \ezTrans\ has mass dimension $n$, the typical value of the potential goes like $k_i^2 \Lambda^{2n-2}$(each $x_i$ might have different dimensions, e.g. $u_r={\rm tr}\phi^r$ for $SU(N)$ case, but they can be made to have the same dimension by multiplying $\Lambda$ appropriately). Then the bounce action $S$ scales like $\Lambda^{6-2n}/k^2$. As long as this quantity is large enough, metastable vacua are long-lived.

\bigskip
\centerline{\bf Acknowledgments}
\bigskip
H.O. thanks the hospitality of the high energy
theory group at the University of Tokyo at Hongo.
This research is supported in part by
DOE grant DE-FG03-92-ER40701.
The research of H.O. is also supported in part
by the NSF grant OISE-0403366 and by the 21st Century COE Program
at the University of Tokyo.
Y.O. is also supported in part by the JSPS Fellowship
for Research Abroad. C.P. is also supported in part by
Samsung Scholarship.

%%%%%%%%%%%%%%%%%%%%%%%%%%%%%%%%%%%%%%%%%%%%%%%%%%%%%%%%%%%%%%%%%%%%%%%%%%%%%
\appendix{A}{Semi-classical consideration}
%%%%%%%%%%%%%%%%%%%%%%%%%%%%%%%%%%%%%%%%%%%%%%%%%%%%%%%%%%%%%%%%%%%%%%%%%%%%%
Although it is very complicated to derive explicit geometric quantities for general points of the moduli space of ${\cal N}=2$ $SU(N)$ supersymmetric gauge theory, there are simple expressions available in the semi-classical region. This is the region where only perturbative corrections are enough. For simplicity, we consider the case without hypermultiplets. Since there is a possibility that the curvature is semi-positive and the flat directions are not lifted for any choice of $k_i$ in $W=k_i z^i$, it is useful to see that this actually does not happen in this regime. In the semi-classical approximation, we have to consider the region in which each $A_i$ in \eLeffSemi\ is different from each other to prevent enhanced gauge symmetry. The prepotential ${\cal F}(A)$ in \eLeffSemi\ is given by, up to nonperturbative corrections \SeibergUR ,
$${ {\cal F}(A)={N \tau_0\over 2} \sum_i \left(A_i - {\sum_j A_j\over N} \right)^2+ {i\over 4\pi} \sum_{i<j} (A_i-A_j)^2\log {(A_i-A_j)^2\over \Lambda^2}\;.}$$
Here $A_i$ are the coordinates of a point $p$ in the moduli space. Of course, they are not independent and subject to the constraint $\sum_i A_i =0$. From this,
it is straightforward to derive the various metric and their derivative components. In particular, the curvature does not vanish and is of order $O(g^4)$. Hence we see that the nonzero curvature is induced by the perturbative effects. The derivatives of the metric are given by
$${\partial_k g_{j\qb} = {1\over 2\pi} \left(\delta_{j\qb}\delta_{jk} \sum_m {1\over A_j -A_m} - {\delta_{j\qb}\over A_j-A_k} -{\delta_{jk}-\delta_{\qb k}\over A_j - A_{\qb}}\right)\;.}$$

If we contract this with a vector $w^j$ at $p$,
\eqn\ePSC{{P_{k\qb}}=w^j \partial_k g_{j\qb}={1\over 2\pi} \sum_m \left({w^k-w^m\over A_k-A_m}\right) \delta_{k\qb} - {w^k-w^{\qb}\over A_k-A_{\qb}}\;,}
where we implicitly omit terms whose denominators vanish. Note that this is precisely the expression that entered \ePositiveDefinite . $P_{k\qb}$ in \ePSC , treated as a matrix, is nonsingular at least at one value of $w^j$: When $w^j=A_j$,
$${P_{k\qb}={N\over 2\pi} \left( \delta_{k\qb} - {1\over N}\right)\;,}$$
which is non-degenerate (note that the vector $(1,\cdots,1)$ does not count). This implies $P_{k\qb}$ is non-degenerate for generic choices of $w^j$. In \ePositiveDefinite , $g^{\qb p}$ is positive definite. So the equality holds only when $v=0$ for the above given $w$. Therefore, we can choose a superpotential to make a metastable vacuum at any point in the semi-classical regime.

%%%%%%%%%%%%%%%%%%%%%%%%%%%%%%%%%%%%%%%%%%%%%%%%%%%%%%%%%%%%%%%%%%%%%%%%%%%%%
\appendix{B}{Metastable vacua at the origin of the $SU(N)$ moduli space}
%%%%%%%%%%%%%%%%%%%%%%%%%%%%%%%%%%%%%%%%%%%%%%%%%%%%%%%%%%%%%%%%%%%%%%%%%%%%%

In section 2, we showed how our mechanism applied to the simplest case when
the gauge group is $SU(2)$.
We can extend this to the more general $SU(N)$.
For simplicity, we will consider the $SU(N)$ theory without
hypermultiplet. Though it is hard to find an explicit form of the moduli space metric for
the $SU(N)$ theory and compute its curvature, it turns out to be possible at the origin of
the moduli space. This result in turn determines the normal coordinates and hence
the superpotential which generates a metastable vacuum at the origin.
Later, we consider a deformation of the superpotential so that it becomes a single-trace
operator.

Let $u_r={\rm tr} (\phi^r)$, $i=1,\cdots, N$. These parameterize the moduli space. They become $u_r=\sum_i (a_i)^r$ at weak coupling where $a_i$ are the expectation values of the eigenvalues of the chiral supermultiplet. It is more convenient to use the symmetric polynomials whose expressions at weak coupling are given by
$${
s_r=(-1)^r \sum_{i_1 < \cdots < i_r} a_{i_1}\cdots a_{i_r},\qquad r=2,\cdots,N\;.
}$$
At strong coupling, these are defined by
\eqn\eNewton{
r s_r + \sum^r_{\alpha=0} s_{r-\alpha} u_\alpha = 0,\qquad r=1,2,\cdots\;.
}
The moduli space are given by the elliptic curve \refs{\KlemmQS,\ArgyresXH}:
\eqn\eEllipticCurve{
y^2=P(x)^2 - \Lambda^2 \qquad {\rm where}\qquad P(x)=\sum^N_{\alpha=0} s_{\alpha} x^{N-\alpha}\;.
}
At the origin of the moduli space, all $s_r=0$ and $P(x)=x^N$. $s_0$ is defined to be 1 and $s_1=0$ for $SU(N)$ case.

We choose the basis cycles $\alpha_i$ and $\beta_j$ such that their intersection form is $(\alpha_i, \beta_j)=\delta_{ij}$, $i,j=1,\cdots, N-1$. Then
\eqn\eAAD{
a_{Di}=\oint_{\alpha_i} \lambda,\qquad a_{j}=\oint_{\beta_j} \lambda
}
where
$${
\lambda = \sum^{N-1}_{\alpha=0} (N-\alpha) s_\alpha x^{N-\alpha} {dx\over y}\;.
}$$
There is an overall constant in front of $\lambda$ which can be determined by examining the classical limit. But it can be absorbed in the coefficients $k_\alpha$ in the superpotential $W=k_\alpha z^\alpha$. So the exact coefficient is not necessary.
Since
$${
{\partial \lambda \over \partial s_{\alpha}}= -{x^{N-\alpha}\over y} + d\left( {x^{N+1-\alpha} \over y} \right)\;,
}$$
the differentials of $a_D$ and $a$ are

\eqn\eDAAD{
{\partial a_{Di} \over \partial s_{\alpha}} =- \oint_{\alpha_i}{x^{N-\alpha}\over y} ,\qquad {\partial a_{j} \over \partial s_{\alpha}} =- \oint_{\beta_j}{x^{N-\alpha}\over y}\;.
}
Since we are going to compute the connection and curvature at the origin, we also need expressions for multiple differentiation. Differentiating the above equation with respect to $s_{\beta}$,
$${\eqalign{
{\partial^2 \lambda \over \partial s_{\alpha}\partial s_{\beta}} &\simeq {x^{N-\alpha}\over y^3} P(x) x^{N-\beta} dx\cr
&=\sum_{\rho=0}^N s_{\rho} {x^{3N-\alpha-\beta-\rho}\over y^3} dx\;,
}}$$
where $\simeq$ means equality up to exact pieces.
Differentiating once more,
$${
{\partial^3 \lambda \over \partial s_{\alpha}\partial s_{\beta}\partial s_{\gamma}}\simeq\sum_{\rho=0}^N s_{\rho} {-3x^{4N-\alpha-\beta-\gamma-\rho} \over y^5} P(x) dx + {x^{3N-\alpha-\beta-\gamma}\over y^3 } dx \;.
}$$
These are general expressions. Now we consider the values at the origin of the moduli space.
Using the relations
$${\eqalign{
d\left({x^{N-k} \over y} \right) &= (N-k){x^{N-k-1}\over y} dx - {N x^{3N-k-1} \over y^3} dx\cr
d\left({x^{3N-k} \over y^3} \right) &= (3N-k){x^{3N-k-1}\over y^3} dx - {3N x^{5N-k-1}\over y^5} dx\;,
}}$$
it follows that
\eqn\eDDAAD{\eqalign{
{\partial^2 \lambda \over \partial s_{\alpha}\partial s_{\beta}} &\simeq {N-\alpha-\beta+1\over N} {x^{N-\alpha-\beta} \over y } dx\cr
{\partial^3 \lambda \over \partial s_{\alpha}\partial s_{\beta}\partial s_{\gamma}} &\simeq {(\alpha+\beta+\gamma-2N-1)(N-\alpha-\beta-\gamma+1) \over N^2} {x^{N-\alpha-\beta-\gamma} \over y } dx\;.
}}

When the moduli are set to the origin, the curve is given by $y^2=x^{2N}-1$. Here we set the scale $\Lambda$ of the theory to 1. We place the branches on the unit circle as follows
\refs{\DouglasNW, \AlberghiTU} : The $n$-th branch lies along the angle ${2\pi\over N}(2n-2)$ to ${2\pi\over N} (2n-1)$. The $\alpha_n$ cycle encloses the $n$-th branch. The $\gamma_n$ cycle runs between $n-1$ and $n$-th branches(indices are modulo n). For example, when $N=4$, the branches are distributed as in Figure 3.

\bigskip
\centerline{\epsfxsize=0.4\hsize\epsfbox{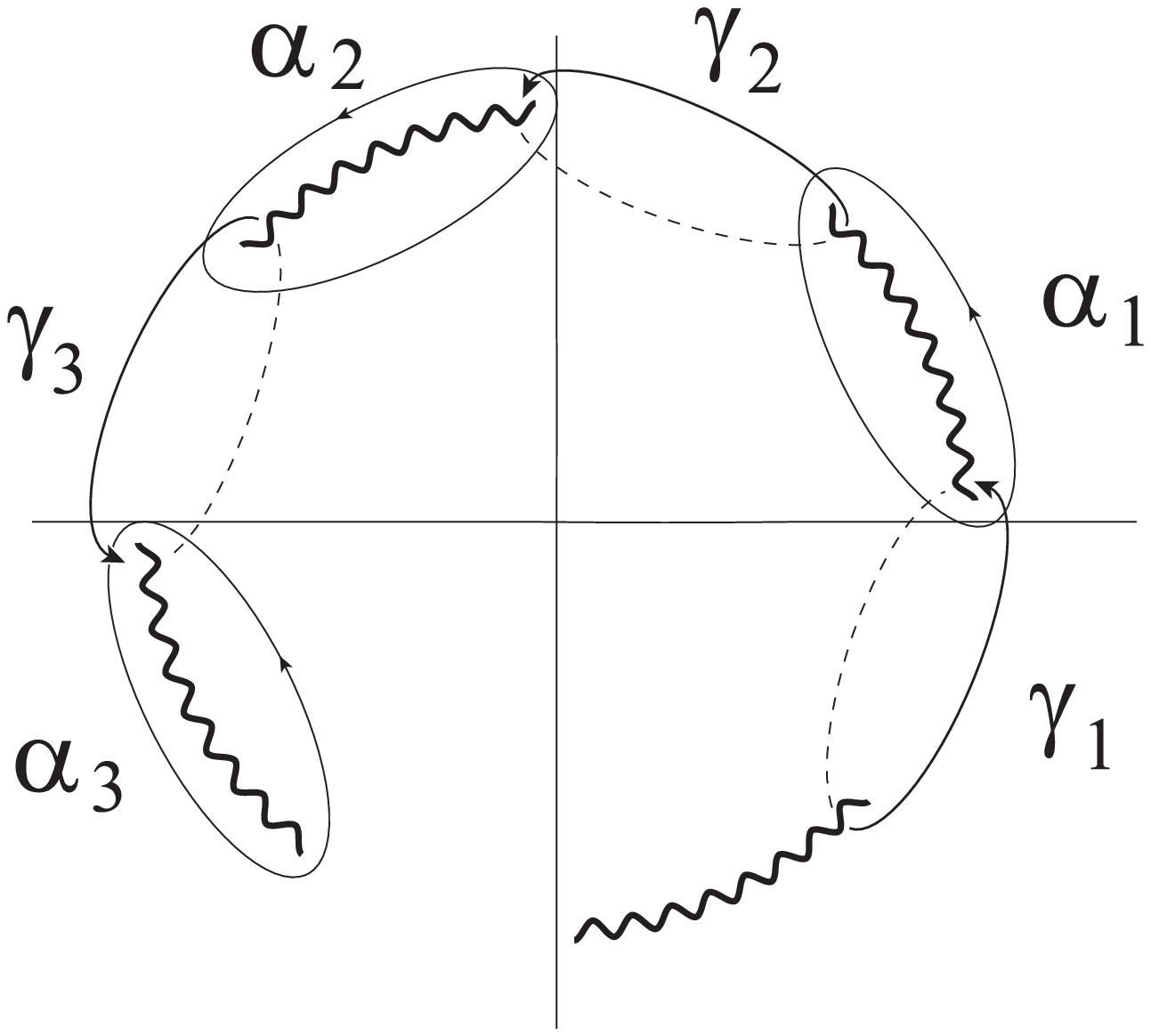}}
\noindent{\ninepoint\sl \baselineskip=8pt {\bf Figure 3}:{\sl
The cycles $\alpha_i$ and $\gamma_i$, and the branches for the moduli $s_\alpha=0$ when $N=4$.}}
\bigskip

We choose the cycles $\beta_n$ by
\eqn\eBetaCycle{
\beta_n=\sum_{i\le n} \gamma_i\;.
}
Then the intersection matrix for $\alpha_m$ and $\beta_n$ are given by $(\alpha_m, \beta_n)=\delta_{mn}$. Since we are considering the moduli space at the origin, the periods have many relations among each other. These eventually determine all periods in terms of one function. Let's start with the period
$${\eqalign{
{\partial a_{Dm}\over \partial s_{\alpha}} &= -\oint_{\alpha_m} {x^{N-\alpha}\over y} dx \cr
&=-2 \int_{2\pi (m-1)/N}^{2\pi(m-{1\over2})/N} {e^{i \theta(N-\alpha)} e^{i\theta} \over \sqrt{e^{2 i N\theta}-1}} i d\theta
\;.}}$$
By changing integration variables, we get the recursion relation
$${
{\partial a_{Dm+1}\over \partial s_{\alpha}}=e^{{2\pi i\over N} (N-\alpha+1)}{\partial a_{Dm}\over \partial s_{\alpha}}\;.
}$$
That is,
$${
{\partial a_{Dm}\over \partial s_{\alpha}}=e^{{2\pi i\over N} (N-\alpha+1)(m-1)}{\partial a_{D1}\over \partial s_{\alpha}}\;,
}$$
and the same relations hold for their differentiations with respect to $s_\beta$(resp. $s_\beta$ and $s_\gamma$) by replacing $\alpha$ with $\alpha+\beta$(resp. $\alpha+\beta+\gamma$). Also, an analogous result can be drawn for $a$ by using the cycle $\beta_n$ in \eBetaCycle\ .
Moreover, $a_{D1}$ and $a_1$ are related by
$${\eqalign{
{\partial a_{1}\over \partial s_{\alpha}}&=e^{-{i\pi\over N} (N-\alpha+1)} {\partial a_{D1}\over \partial s_\alpha}\cr
{\partial^2 a_{1}\over \partial s_{\alpha}\partial s_\beta}&=e^{-i{\pi\over N} (N-\alpha-\beta+1)} {\partial^2 a_{D1}\over \partial s_\alpha\partial s_\beta}\cr
{\partial^3 a_{1}\over \partial s_{\alpha}\partial s_\beta \partial s_\gamma}&=e^{-i{\pi\over N}(N-\alpha-\beta-\gamma+1)} {\partial^3 a_{D1}\over \partial s_\alpha\partial s_\beta \partial s_\gamma}\;,
}}$$
which can also be obtained by change of integration variables. So, let us define
$${
h(\alpha)=2\int_0^{\pi/N} {e^{i\theta(N-\alpha+1)} \over \sqrt{1-e^{2N i \theta}}} d\theta
}$$
so that
$${
{\partial a_{D1}\over \partial s_\alpha}= -h(\alpha)\;.
}$$
The rest are determined by the above relations.

The metric is given by
$${
g_{\alpha\beb} = {1\over 2 i} \sum_j \left( {\partial a_{Dj}\over \partial s_\alpha}{\partial \ab_{j}\over \partial \sb_\beb}-{\partial a_{j}\over \partial s_\alpha}{\partial \ab_{Dj}\over \partial \sb_\beb}\right)\;.
}$$
By substitution, we reach
$${
g_{\alpha\beb}=A_{\alpha,\beb} h(\alpha)\bar{h}(\beb)\;,
}$$
where
$${\eqalign{
A_{\alpha,\beb}=&{1\over 2 i}\sum_{i=1}^{N-1}\sum_{j=1}^i e^{{2\pi i\over N}\left[(N-\alpha+1)(i-1) -(N-\beb+1)(j-1)+{1\over 2} (N-\beb+1)\right]}\cr
&-{1\over 2 i}\sum_{i=1}^{N-1}\sum_{j=1}^i e^{{2\pi i\over N}\left[(N-\alpha+1)(j-1) -(N-\beb+1)(i-1)-{1\over 2} (N-\alpha+1)\right]}\;.
}}$$
The summation can be done straightforwardly. This is nonzero only when $\alpha=\beb$ provided $\alpha,\beb \le N$. Evaluating when $\alpha=\beb$, we get
$${
A_{\alpha,\beb}={N\over 2 \sin{\pi (\beb-1)\over N}} \delta_{\alpha,\beb}\;.
}$$
When evaluating $\partial_\gamma g_{\alpha\beb}$, we get a very similar expression but with $A_{\alpha+\gamma,\beb}$ instead of $A_{\alpha,\beb}$. Since $\alpha+\gamma$ can be $N+1$, in which case $A_{\alpha+\gamma,\beb}$ is non-zero, it may cause a problem. But, fortunately, such terms do not contribute by \eDDAAD . $A_{\rho+\gamma+\alpha,\beb}$ is nonzero when $\rho+\gamma+\alpha=\beb+2N$ and we have to take this into account.

The results of the computation are summarized as follows:
\eqn\eSumGCC{\eqalign{
g_{\alpha\beb}=&\delta_{\alpha\beb} {N\over 2 \sin{\pi (\beb-1)\over N}} |h(\beb)|^2\cr
g^{\alpha\beb}=&\delta_{\alpha\beb} {2 \sin{\pi (\beb-1)\over N}\over N} |h(\beb)|^{-2}\cr
\partial_\gamma g_{\alpha\beb} =& -\delta_{\gamma+\alpha,\beb} {N-\beb+1\over 2 \sin{\pi (\beb-1)\over N}}|h(\beb)|^2\cr
\partial_\rho \partial_\gamma g_{\alpha\beb} =& -\delta_{\rho+\gamma+\alpha,\beb} {(N-\beb+1)(\beb-2N-1)\over 2 N \sin{\pi (\beb-1)\over N}}|h(\beb)|^2\cr
&+\delta_{\rho+\gamma+\alpha,\beb+2N}{(\beb-1)^2\over 2 N \sin{\pi (\beb-1)\over N}}|h(\beb)|^2\cr
\partial_{\deb}\partial_{\gamma}g_{\beta\rhb}=&\delta_{\beta+\gamma,\rhb+\deb} {(N-\beta-\gamma+1)^2 \over 2 N \sin {\pi(\beta+\gamma-1)\over N}}|h(\beta+\gamma)|^2\cr
\Gamma^{\alpha}_{\,\,\beta\gamma}=&g^{\alpha\deb}\partial_\beta g_{\gamma\deb}= - \delta_{\alpha,\beta+\gamma}{N-\alpha+1\over N}\cr
R^{\alpha}_{\,\,\beta\gamma\deb}=&g^{\alpha \pb} g^{q\rhb} \partial_{\deb}g_{q\pb}\partial_{\gamma}g_{\beta\rhb}-g^{\alpha\rhb}\partial_{\deb}\partial_{\gamma}g_{\beta\rhb}\cr
=&\cases{-\delta_{\alpha+\deb,\beta+\gamma} {(N-\alpha-\deb+1)^2\over N^2} \left| {h(\alpha+\deb)\over h(\alpha)}\right|^2 {\sin{\pi (\alpha-1)\over N}\over \sin{\pi(\alpha+\deb-1)\over N}} & for $\alpha+\deb>N$\cr
0 & otherwise}\;.
}}

Let us try $W=\lambda s_\alpha$ as our starting superpotential where $\lambda$ is a small coupling constant. Due to the curvature formula, $R^{\alpha}_{\,\,\alpha \beta \gab}$ for fixed $\alpha$ is a diagonal matrix with some zeroes on the diagonal unless $\alpha=N$. Hence the only plausible case is $W=\lambda s_N$. In this case,
\eqn\eRuddd{
R^N_{\,\,N\gamma\deb}=\delta_{\gamma\deb}{(\deb-1)^2\over N^2} \left| {h(N+\deb)\over h(N)} \right|^2 {\sin{\pi\over N}\over\sin{\pi(\deb-1)\over N}}\;,
}
which is manifestly positive-definite.

The correction we need to add to make a normal coordinate is given by \ezTrans . Using the following values
$${\eqalign{
g_{N\Nb}&= {N\over 2\sin{\pi\over N}} |h(N)|^2\cr
\partial_\alpha g_{N-\alpha, \Nb} &= -{1\over 2\sin{\pi\over N}} |h(N)|^2\cr
\partial_\alpha \partial_\beta g_{N-\alpha-\beta, \Nb}&= {1\over 2\sin{\pi\over N}} {N+1\over N}|h(N)|^2\cr
\partial_N \partial_N g_{N, \Nb}&= {1\over 2\sin{\pi\over N}} {(N-1)^2\over N}|h(N)|^2\;,
}}$$
we have
\eqn\eWGeneralN{
W=\lambda z^N=\lambda\left(s_N - {1\over 2N} \sum_{\alpha+\beta=N} s_\alpha s_\beta + {N+1\over 6N^2} \sum_{\alpha+\beta+\gamma=N} s_\alpha s_\beta s_\gamma + {(N-1)^2\over 6N^2} (s_N)^3\right)\;.
}

In the case $N=2$, we have $W=\lambda u + {1\over 24} \lambda u^3,\, u=-s_2$, which is the superpotential that we used to check the metastability for $SU(2)$.

%%%%%%%%%%%%%%%%%%%%%%%%%%%%%%%%%%%%%%%%%%%%%%%%%%%%%%%%%%%%%%%%%%%%%%%%%%%%%
\subsec{Deformation to a superpotential with single-trace terms}
%%%%%%%%%%%%%%%%%%%%%%%%%%%%%%%%%%%%%%%%%%%%%%%%%%%%%%%%%%%%%%%%%%%%%%%%%%%%%

The superpotential \eWGeneralN\ is not a sum of $u_r$ where $u_r={\tr (\phi^r)}$. Actually, $s_\alpha$ is given by the implicit relation \eNewton\ and there are quadratic and cubic terms in $s$ in \eWGeneralN  . For $N=2$ and $3$, the superpotentials are already of single-trace type because we have few independent coordinates ($s_2,\cdots, s_N$). For $N=2$, it is trivial. Let us consider $N=3$. Here, all $\partial_\alpha g_{\beta,\bar{3}}$ vanish since $\alpha+\beta=3$ cannot be satisfied both being greater than or equal to 2. Considering other terms also similarly, the only terms we get are $s_3$ and $(s_3)^3$. $s_3=-u_3/3$ and $(s_3)^3=-u_9/3$ up to cubic orders of $u_2$ and $u_3$. But, for large $N$, this does not work. So we have to consider a deformation.

We will first consider a general deformation and apply this to our case. Given a superpotential $W=k_\alpha z^\alpha$, consider a deformation of the form
\eqn\eWDEF{
W=k_{\alpha} z^{\alpha} + {\alpha_{\alpha\beta}\over 2 } z^{\alpha} z^{\beta} + {\beta_{\alpha\beta\gamma}\over 3} z^{\alpha} z^{\beta} z^{\gamma} \;.
}
We may add quartic or higher degree terms in $z$. This will not change the local behavior of the leading potential near $p$, however. From the inverse metric \eInvm, the leading effective potential is given by
\eqn\eVEFF{\eqalign{ V&=\left(g^{\alpha\deb}+R^{\alpha\deb}_{\,\,\,\,\,\,\, \rho\lab} z^{\rho} \zb^{\lab} \right) \left( k_{\alpha} +\alpha_{\alpha\beta} z^{\beta} + \beta_{\alpha\beta\gamma} z^{\beta} z^{\gamma} \right) \left( \kb_{\deb} +\alb_{\deb\beb}\zb^{\beb} +\beb_{\deb\beb\gab}\zb^{\beb} \zb^{\gab}\right)\cr
&=k_{\alpha} \kb^{\alpha} + \alpha_{\alpha\beta} \kb^{\alpha} z^{\beta}  + \alb_{\alb\beb}k^{\alb}\zb^{\beb}+ \beta_{\alpha\beta\gamma} \kb^{\alpha} z^{\beta} z^{\gamma} + \beb_{\deb\beb\gab} k^{\deb} \zb^{\beb} \zb^{\gab}\cr &\qquad + \left( k_{\rho} \kb_{\lab} R^{\rho\lab}_{\,\,\,\,\,\,\,\beta\gab} + \alpha_{\alpha\beta} \alb^{\alpha}_{\,\,\,\gab} \right) z^{\beta} \zb^{\gab}+O(z^3)\;,}}
where $g^{\alpha\beb}$ and $g_{\alpha\beb}$ are used to raise and lower indices. All tensors are evaluated at the origin. If we demand a deformation leave the local minimum invariant, $\alpha_{\alpha\beta}$ should satisfy
\eqn\eAlpha{ \alpha_{\alpha\beta} \kb^{\beta}=0\;.}
Given such $\alpha_{\alpha\beta}$, \eVEFF\ becomes
\eqn\eVEffS{V=k_{\alpha} \kb^{\alpha}+M_{\alpha\beb} z^{\alpha} \zb^{\beb} + L_{\alpha\beta} z^{\alpha} z^{\beta} + \bar{L}_{\alb\beb} \zb^{\alb} \zb^{\beb}\;,}
where $M_{\alpha\beb}=k_{\rho} \kb_{\deb} R^{\rho\deb}_{\,\,\,\,\, \alpha\beb} + \alpha_{\gamma\alpha}\alb^{\gamma}_{\,\,\beb}$ and $L_{\alpha\beta}= \kb^{\gamma} \beta_{\gamma\alpha\beta}$.
The second term is positive definite, so it tends to give a local minimum at $p$. But the last two terms develop tachyonic directions. So, roughly, when $\beta_{\gamma\alpha\beta}$ is smaller than the order of $k_{\rho} R^{\rho}_{\,\,\gamma\alpha\beta}$ schematically, we have a metastable minimum.

We now consider a specific deformation. Note that the last term of \eWGeneralN\ can be converted into $-{(N-1)^2\over 6N^5} (u_N)^3$ to cubic order and this is
$${
-{(N-1)^2\over 6N^3} u_{3N}
}$$
to the same order.\foot{Actually, the chiral ring is modified due to instantons as discussed in appendix A of \CachazoRY . We will discuss this effect later.} So the last term is fine. That is, if we express $u_{3N}$ in terms of $u_2,\cdots,u_N$, we have a term ${1\over N^2} (u_N)^3$, but all other terms are of quartic and higher orders.

To deform the first three terms of \eWGeneralN , note first that, from \eNewton ,
$${\eqalign{
u_N&=-N s_N - \sum_{\alpha=1}^{N-1} s_{N-\alpha} u_{\alpha}\cr
&=-N s_N - \sum_{\alpha=1}^{N-1} s_{N-\alpha} (-\alpha s_\alpha - \sum_{\beta=1}^{\alpha-1} s_{\alpha-\beta} u_\beta)\cr
&=-N s_N +\sum_{\alpha=1}^{N-1} \alpha s_{N-\alpha} s_\alpha -\sum_{\alpha=1}^{N-1}\sum_{\beta=1}^{\alpha-1} \beta s_{N-\alpha} s_{\alpha-\beta} s_\beta + O(s^4)\;.
}}$$
Therefore,
\eqn\euands{
{1\over N} u_N = - s_N + {1\over 2} \sum_{\alpha+\beta=N} s_\alpha s_\beta - {1\over 3} \sum_{\alpha+\beta+\gamma=N} s_\alpha s_\beta s_\gamma + O(s^4)\;.
}
We can invert \ezTrans\ and get
\eqn\eInvertsz{
s^\rho = z^{\rho} - {1\over 2} g^{\rho \alb}\partial_\beta g_{\gamma\alb} z^\beta z^\gamma +O(z^3)\;.
}
We will consider a superpotential
\eqn\eWSingleTrace{
W=\lambda \left({1\over N} u_N + {(N-1)^2\over 6N^3} u_{3N}\right)
}
for small coupling constant $\lambda$. Note that we have set the scale $\Lambda$ of the theory to 1.

This is indeed a sum of single-trace operators. We will see that this superpotential produces a metastable vacuum at the origin. Note that $(s_N)^3=-{1\over N}u_{3N} + O(s^4)$. Using \eWGeneralN , \euands\ and \eInvertsz , we can express $W$ in terms of $z^\alpha$:
$${\eqalign{
-{\lambda^{-1} W}=&z^N + {1-N\over 2N} \sum_{\alpha+\beta=N} z^\alpha z^\beta \cr
&+ \left({1\over 3} - {N+1\over 6N^2} \right)\sum_{\alpha+\beta+\gamma=N} z^\alpha z^\beta z^\gamma-\sum_{\alpha+\delta=N} g^{\delta \rhb} \partial_\gamma g_{\beta\rhb} z^\alpha z^\beta z^\gamma +O(z^4)\;.
}}$$
Referring to \eWDEF , the deformation corresponds to
\eqn\eABSingleTrace{\eqalign{
-{\lambda^{-1}}\alpha_{\alpha\beta}&= {1-N\over N} \delta_{\alpha+\beta, N}\cr
-{\lambda^{-1}}\beta_{\alpha\beta\gamma} &=  \left(1 - {N+1\over 2N^2} \right) \delta_{\alpha+\beta+\gamma,N} - 3 \sum_{\delta, \rhb} \delta_{N-\delta,(\alpha} g^{\delta \rhb} \partial_\gamma g_{\beta) \rhb}\;,
}}
where $(\cdots)$ in indices denotes symmetrization.

When we deform the superpotential $W$ according to \eWDEF, the tree level potential is given by \eVEFF . From this, we see that deformations given by $\alpha_{\alpha\beta}$ and $\beta_{\alpha\beta\gamma}$ such that
\eqn\eVAR{\eqalign{
\alpha_{\alpha\beta}\kb^\beta &=0\cr
\beta_{\alpha\beta\gamma}\kb^\gamma &=0}}
leave the metastable vacuum at the origin of the effective potential. Since the metric is diagonal at the origin, these amount to requiring $\alpha_{\alpha N}=\beta_{\alpha\beta N}=0$. Note that $g^{\delta\rhb}$ vanish unless $\delta=\rhb$ and $\partial_\gamma g_{\beta \rhb}$ vanish unless $\gamma+\beta=\rhb$. Considering all combinations of indices, $\alpha_{\alpha N}=\beta_{\alpha\beta N}=0$.

As noted before, we also have instanton corrections on the chiral ring. Quantum mechanically\CachazoRY ,
$${
u_{3N} = \sum_{m=0}^{1} \pmatrix{2m \cr m} \Lambda^{2N m} {1\over 2\pi i} \oint_{C} z^{3N} {P'(z) \over P(z)^{2m+1}} dz\;,
}$$
where $C$ is a large contour around $z=\infty$.

The $m=0$ term gives the classical relation, i.e. $u_{3N}={1\over N^2} (u_N)^3+O(u^4)$. The $m=1$ term gives the instanton correction. This changes the coefficients $\alpha_{\alpha\beta}$ and $\beta_{\alpha\beta\gamma}$. However, the relations \eVAR\ are still satisfied since the additional contribution to $\alpha_{\alpha\beta}$(resp. $\beta_{\alpha\beta\gamma}$) occurs only when $\alpha+\beta=N$(resp. $\alpha+\beta+\gamma=N$). We conclude that the superpotential \eWSingleTrace\ gives a metastable vacuum at the origin.

%%%%%%%%%%%%%%%%%%%%%%%%%%%%%%%%%%%%%%%%%%%%%%%%%%%%%%%%%%%%%%%%%%%%%%%%%%%%%
\subsec{Large $N$ behavior}
%%%%%%%%%%%%%%%%%%%%%%%%%%%%%%%%%%%%%%%%%%%%%%%%%%%%%%%%%%%%%%%%%%%%%%%%%%%%%
The first and the second terms of \eWSingleTrace\ are both of order $N^{-1}$ when expressed in terms of $s_\alpha$. Hence we may consider a deformation that eliminates the second term. This turns out not to be possible.

Since $\lambda$ is just an overall coefficient, we can set it to -1 in the following discussion. Note that $u_{3N}$ is $-N (z^N)^3$ in $z^\alpha$ coordinates to cubic order. Hence \eABSingleTrace\ change to
\eqn\eABSingleTrace{\eqalign{
\alpha_{\alpha\beta}&= {1-N\over N} \delta_{\alpha+\beta, N}\cr
\beta_{\alpha\beta\gamma} &= \left(1 - {N+1\over 2N^2} \right)  \delta_{\alpha+\beta+\gamma,N} - 3\delta_{N-\delta,(\alpha} g^{\delta \rhb} \partial_\gamma g_{\beta) \rhb}-{(N-1)^2\over 2N^2} \delta_{\alpha,N}\delta_{\beta,N}\delta_{\gamma,N}\;.
}}
Since we have shown that $\alpha_{\alpha N}=\beta_{\alpha\beta N}=0$ were it not for the additional term, we have
$${
\beta_{\alpha\beta N}=-{(N-1)^2\over 2N^2} \delta_{\alpha,N}\delta_{\beta,N}\;.
}$$
Then $L_{\alpha\beta}=g^{N\Nb}\beta_{\alpha\beta N}$ in \eVEffS\ are all zero except when $\alpha=\beta=N$ and
$${
L_{NN}=-{(N-1)^2\over N^3} {\sin{\pi \over N}\over |h(N)|^2}\;.
}$$
Since $h(N)\sim 1/N$, this scales like $N^0$ for large $N$. But $M_{\alpha\beb}$ are given by, using \eRuddd ,
$${\eqalign{
M_{\alpha\beb}&=R^{N\Nb}_{\,\,\,\,\,\,\,\,\alpha\beb} +\alpha_{\gamma\alpha}\alb^{\gamma}_{\,\,\beb}=g^{N\Nb} R^{N}_{\,\,\,N\alpha\beb}+ g^{\gamma\deb} \alpha_{\gamma\alpha} \alb_{\deb\beb}\cr
&=\delta_{\alpha\beb} {2(\beb-1)^2\over N^3} {\left| h(N+\beb)\right|^2 \over \left|h(N)\right|^4} {(\sin{\pi\over N})^2\over\sin{\pi(\beb-1)\over N}}\cr
&\qquad\qquad\qquad+\left({1-N\over N}\right)^2 \sum_{\gamma} {2\sin{\pi(\gamma-1)\over N}\over N} \left| h(\gamma)\right|^{-2} \delta_{\alpha+\gamma, N}\delta_{\beb+\gamma, N}\;.
}}$$
$M_{\alpha\beb}$ are diagonal and the second term vanishes when $\alpha=\beb=N$. Therefore,
$${
M_{N\Nb}={2(N-1)^2\over N^3}{\left| h(2N)\right|^2 \over \left|h(N)\right|^4} \sin{\pi\over N}\;.
}$$
Since $h(2N)$ scales like $N^{-2}$, $M_{N\Nb}$ scales like $N^{-2}$. Since $L_{NN}$ introduces saddle point behavior at the origin along the $N$-th direction and $M_{N\Nb}$ is not large enough to lift it, the metastability could not be maintained in the large $N$ limit if we did not include the second term of \eWSingleTrace . Actually, $L_{NN}/M_{N\Nb} > 1/2$ for all $N$(we have explicit formulae), so we cannot remove the second term of \eWSingleTrace\ for any $N$.

The components of \eRuddd\ is of order $N^{-2}$ when $\gamma=\delta$ is near $N$ and of order $N^{-1}$ when $\gamma=\delta$ is near $N/2$. So although the metastable vacuum at the origin persists for any finite $N$, the mechanism to make metastable vacua using the curvature becomes harder and harder to implement as $N$ increases in the current setup.

\listrefs

\end